# Optimized reversible BCD adder using new reversible logic gates

H.R.Bhagyalakshmi, M.K.Venkatesha

**Abstract**—Reversible logic has received great attention in the recent years due to their ability to reduce the power dissipation which is the main requirement in low power digital design. It has wide applications advanced computing, low power CMOS design, Optical information processing, DNA computing, bio information, quantum computation and nanotechnology. This paper presents an optimized reversible BCD adder using a new reversible gate. A comparative result is presented which shows that the proposed design is more optimized in terms of number of gates, number of garbage outputs and quantum cost than the existing designs.

## 1 INTRODUCTION

Conventional Combinational logic circuits dissipate heat for every bit of information that is lost during their operation. Due to this fact the information once lost cannot be recovered in any way. But the same circuit if it is constructed using the reversible logic gates will allow the recovery of the information. In 1960s R.Landauer demonstrated that even with high technology circuits and systems constructed using irreversible hardware results in energy dissipation due to information loss [1].It is proved that the loss of one bit of information dissipates KTln2 joules of energy where K is the Boltzman's constant and T is the absolute temperature at which the operation is performed[1]. Later Bennett, in 1973, showed that in order to avoid KTln2 joules of energy dissipation in a circuit it must be built from reversible circuits [2].

A reversible logic gate is an n-input, n-output logic device with one-to-one mapping. This helps to determine the outputs from the inputs but also the inputs can be uniquely recovered from the outputs.Extra inputs or outputs are added so that the number of inputs is made equal to the number of outputs whenever it is necessary. An important constraint present on the design of a reversible logic circuit using reversible logic gate is that fan-out is not allowed. A reversible circuit should be designed using minimum number of reversible gates. One key requirement to achieve optimization is that the designed circuit must produce minimum number of garbage outputs.Also they must use minimum number of constant inputs [3, 4].

The present work proposes a BCD adder which uses a new reversible logic gate called SCL gate in combination with the existing reversible logic gates for further improvement of the optimization parameters.

This paper is organized as follows:

Section 1 gives the necessary introduction on reversible logic gates along with the important design issues is discussed. The logic diagrams are given along with their logic function representations instead of the truth tables. Section 2 gives some of the important reversible logic gates along with their logic diagrams and logic functions. The new reversible logic gate called SCL gate designed to get the overflow detection in BCD adder along with its logic function representation is presented. In Section 3 an irreversible BCD adder and the design issues involved is discussed. Also the proposed BCD adder using the existing reversible logic gates and new reversible gate is described. Section 4 gives the result of comparison of the proposed design with the other important existing circuits available in the literature in terms of the parameters of optimization. Section 5 gives the conclusions and future scope of the present work.

- *H.R.Bhagyalakshmi is with the Department of Electronics and communication engineering , B.M.S College of Engineering, Bangalore, India.*
- *M.K.Venkatesha is with the Department of Electronics and communication engineering,R.N.S Institute of technology, Bangalore, India.*

## 2 BASIC REVERSIBLE LOGIC GATES

### 2.1 Reversible logic gate

It is an n-input n-output logic function in which there is a one-to-one correspondence between the inputs and the outputs. Because of this bijective mapping the input vector can be uniquely determined from the output vector. This prevents the loss of information which is the root cause of power dissipation in irreversible logic circuits.

The reversible logic circuits must be constructed under two main constraints. They are

- Fan-out is not permitted.
- Loops or feedbacks are not permitted.

In the proposed design these two constraints along with the other parameters are optimized effectively.

## 2.2 Basic reversible logic gates

The important basic reversible logic gates are, Feynman gate [6] which is the only 2*2 reversible gate which is as shown in the figure.2a and it is used most popularly by the designers for fan-out purposes. There is also a double Feynman gate [7], Fredkin gate [8] and Toffoli gate [9], New Gate[10] , Peres gate[11] , all of which can be used to realize important combinational functions and all are 3*3 reversible gates and are as shown in the figure.2b to figure.2e .The figures also shows the switching functions for terminals.

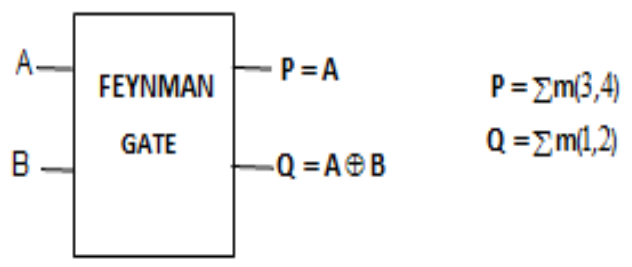

Fig.2a Feynman gate – 2*2 gate

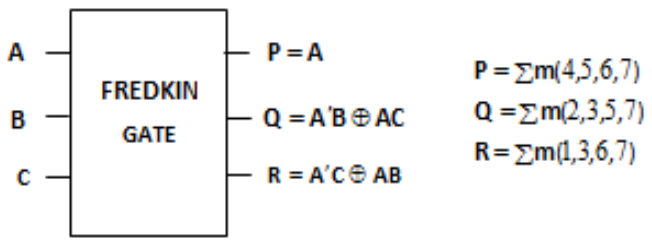

Fig.2b Fredkin gate – 3*3 gate

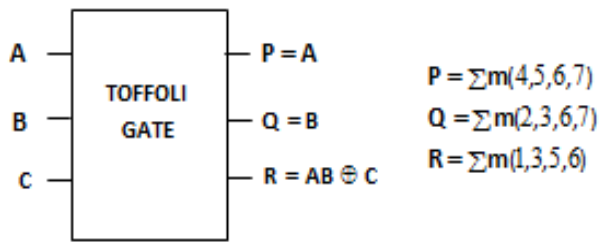

Fig.2c Toffoli gate – 3 * 3 gate

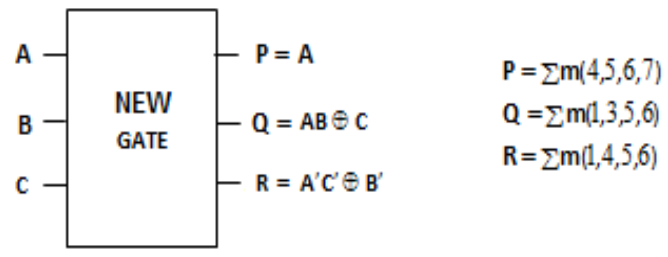

Fig.2d. New gate – 3 * 3 gate

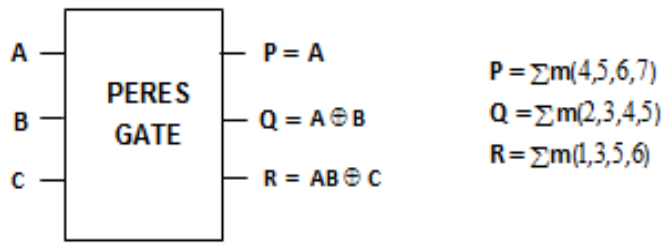

Fig.2e. Peres gate – 3 * 3 gate

There are other 4*4 gates some of which are specially designed for the realization of important combinational circuit functions in addition to some basic functions. Some of the important 4*4 gates are, TSG gate [13], MKG gate [12] ,HNG gate [14] etc, shown in figure 3, all of which are very useful for the construction of important reversible adders. They are also technology independent as quantum logic and optical logic and DNA logic are all still in the initial implementation stages and the technology is not defined properly. However the design methods using existing reversible logic gates and new reversible logic gates are very important and useful for building future computation circuits and also for the design of ultra low power integrated circuits. They are classified as an important theoretical computational circuits.

## 2.3 Optimization parameters

The important parameters which play a major role in the design of an optimized reversible logic circuit are [3-5],

- **Constants:** This refers to the number of inputs that are to be maintained constant at either 0 or 1 in order to synthesize the

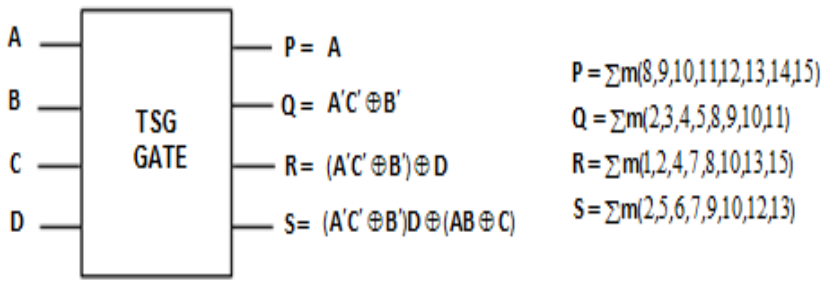

Fig.3a TSG gate – 4 * 4 gate

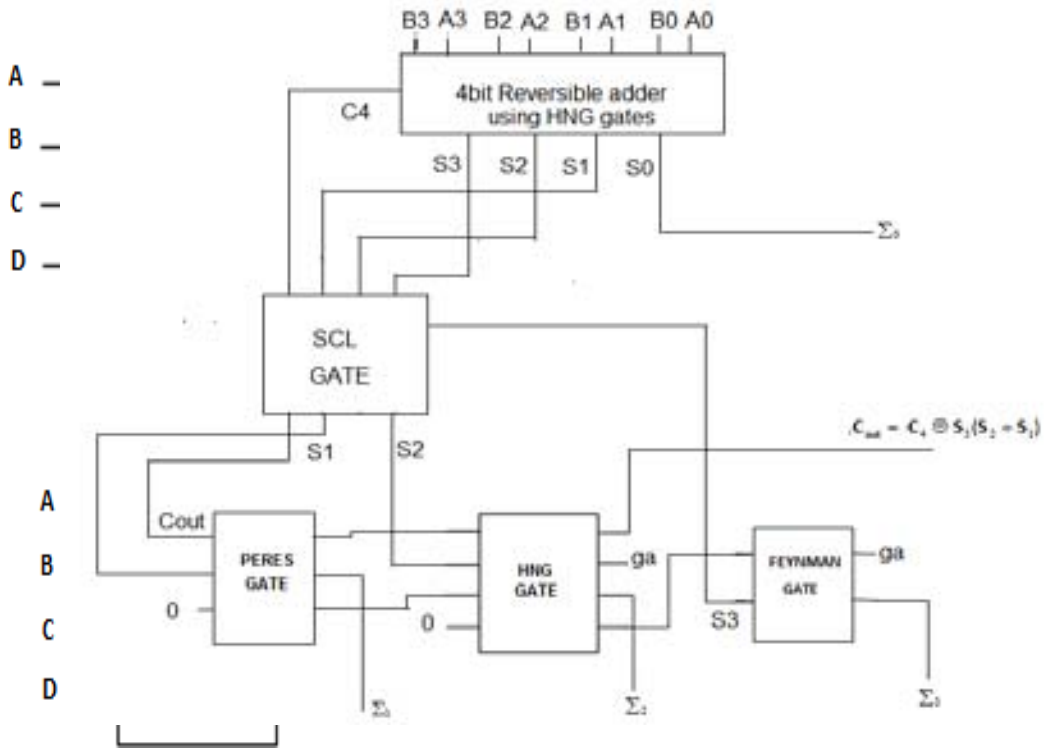

Fig. 7. Reversible implementation of a one digit BCD adder.
Fig.3c. HNG gate – 4 * 4 gate

  given logical function.
- **Garbages:** This refers to the number of outputs which are not used in the synthesis of a given function.These are very essential without which reversibility cannot be achieved.
- **Gate count:** The number of reversible gates used to realize the function.
- **Flexibility:** This refers to the universality of a reversible logic gate in realising more functions.
- **Quantum cost:** This refers to the cost of the circuit in terms of the cost of a primitive gate. It is calculated knowing the number of primitive reversible logic gates (1*1 or 2*2 ) required to realize the circuit
- **Gate levels:** This refers to the number of levels in the circuit which are required to realize the given logic functions.

The present paper proposes one new gate, called SCL gate (Six Correction Logic) for the correction in the BCD addition and is as shown in the Fig 4.It is a 4 * 4 reversible and its logic function representation is as shown.

## 2.4 BCD adder and the design issues

A one digit BCD adder adds two BCD numbers and produces the BCD sum after the required correction which is according to the rules for BCD addition.

The circuit of a conventional irreversible BCD adder is as shown in fig.5.

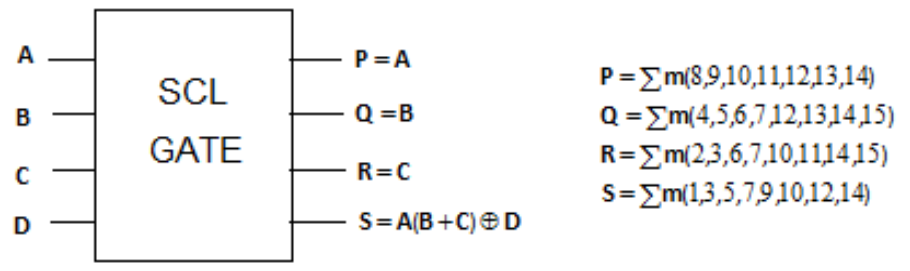

Fig.4 SCL gate –for BCD adder's six-correction logic gate.

## 3 CONSTRUCTION OF BCD ADDER

In a BCD adder, the correction logic which generates the $C_{out}$ is given by,

$$C_{out} = S_3S_2 + S_3S_1 + C_4 \quad .................... \quad \text{(Eq1)}$$

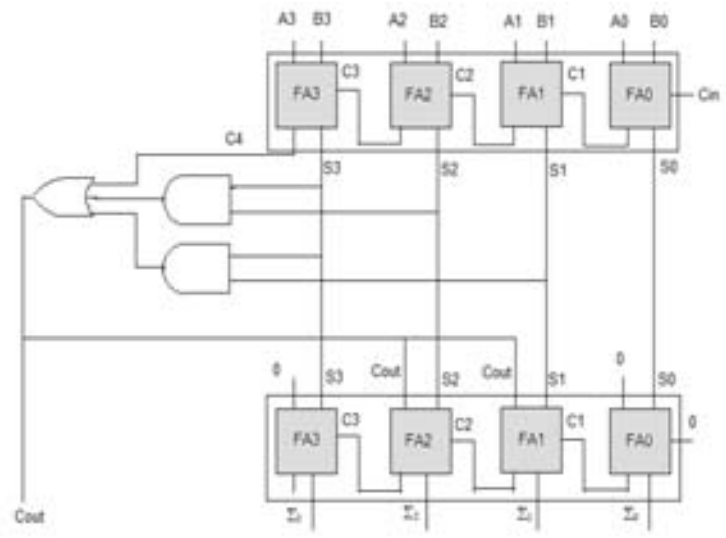

Fig. 5. Conventional 1 digit BCD adder.

The above equation can also be expressed without changing its functionality into,

$$C_{out} = C_4 \oplus S_3 (S_2 + S_1) \quad ..................... \quad \text{(Eq2)}$$

The BCD adder can be constructed using reversible gates. Fig 6 shows the 4 bit parallel adder constructed using HNG gates which can also be constructed using TSG or MKG gates.

The proposed BCD adder circuit uses one such 4 bit parallel adder and is called as adder-1 in this proposal. The total number of garbage outputs generated from the reversible parallel adder is equal to eight. The overflow detection uses one SCL gate. This does not produce any garbage outputs. Also the second adder which should add six in order to correct and convert the sum to BCD sum need not be a 4bit parallel adder but instead it can be constructed using one Peres gate, one HNG gate and one Feynman gate similar to the existing designs [16-17].

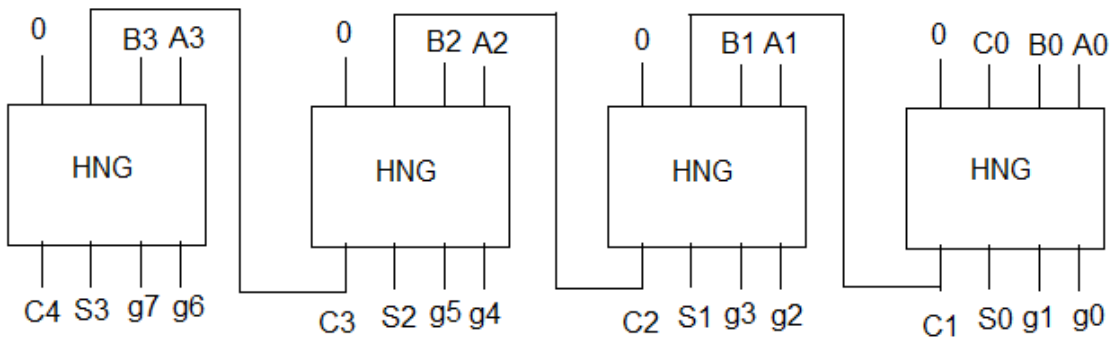

Fig.6. Used reversible 4bit parallel adder

The New gate is used to add S1 with Cout to produce final $\mathbf{\Sigma}_1$ and a carry which is given to one HNG gate used as a full adder to produce final $\mathbf{\Sigma}_2$. Then the final sum bit $\mathbf{\Sigma}_3$ is obtained by using one Feynman gate. So the BCD sum is $\mathbf{\Sigma}_3\mathbf{\Sigma}_2\mathbf{\Sigma}_1\mathbf{\Sigma}_0$.The complete BCD adder is as shown in the fig.7.

## 4 RESULTS AND DISCUSSION

Several researchers have proposed the 4bit parallel adder which is constructed using 4*4 reversible full adder gates [13-17]. It is also known that the full adder circuit requires a minimum of two garbage outputs and a constant input [15].

In [13] 3 New gate for the correction logic circuit and 8 TSG gates for the adders are used for the construction of reversible implementation of BCD adder. This reduces the number of gates but in this paper fan-out is not taken into account which when considered will increase the number of gates above 11. This produces 22 garbage outputs with 11 constant inputs.

In [14] the BCD adder is realized using 8 HNGs along with one HNFG and one FG for fan-outs. It also uses 2 NGs, one TG and one FG for the implementation of the correction logic. This uses a total of 14 reversible gates and it produces 22 garbage outputs with 17 constant inputs in the complete circuit.

In [15] the implementation uses three New gates and four Feynman gates for the correction logic circuit in addition to the eight New gates and eight New Toffoli gates used for the adder circuits. It uses 23 reversible gates producing 22 garbage outputs with 17 constant inputs.

The implementation given in [16] uses two Fredkin gates and one Toffoli gate for the correction logic and also it uses a 4bit parallel adder constructed using four MTSGs. It also uses a combination of one FG, one PG and a MTSG for the adder which adds the cout to the sum in order to generate the final BCD sum. This implementation produces a total number of 11 garbage outputs with 7 constant inputs.

In the implementation of [17] a total of nine gates are used which produces 11 garbage outputs with 7 constant inputs.

In the present paper the design is such that these parameters are kept to the minimum value. Also the full adder uses the existing design which produces two garbage outputs with one constant input. However using the SCL gate for implementing the correction logic of the BCD adder the garbage output of the correction circuit is reduced to zero instead of the minimum one from the existing literature [17].This design is more optimized since the total number of garbage outputs is 10 with only 6 constant inputs.

The proposed circuit uses a total number of 8 reversible gates consisting of five HNG gates, one Peres gate, one Feynman gate and one SCL gate. This is less than the number of gates implemented in [17] and definitely less than the number of gates in the other designs given in [13-16]. Also the number of garbage outputs in the proposed design is 10 which is once again less as compared to that of the designs presented in [13-17].

The total delay of the BCD adder is calculated in terms of the gate delays. If the delay taken to produce the final BCD sum is $\eta_{sum}$ then for a single BCD adder block the total delay is given by, $\eta_{sum} = \eta_{adder1} + \eta_{correction} + \eta_{adder2}$

where
$\eta_{adder1}$ = total delay in the 4bit reversible parallel adder.
$\eta_{correction}$ = delay in generating the Cout.
$\eta_{adder2}$ = delay in generating the final BCD sum.
From the implementation it can be seen that
$\eta_{adder1}$ = 4 HNGs, $\eta_{correction}$ = 1 SCLG
$\eta_{adder2}$ = 1PG + 1 HNG + 1 FG.
Therefore $\eta_{sum}$= 8 gate delays.

## 5 CONCLUSIONS AND FUTURE WORK

In this paper an optimized reversible BCD adder is presented. The design is very useful for the future computing techniques like ultra low power digital circuits and quantum computers. It is shown that the proposal is highly optimized in terms of number of reversible logic gates, number of garbage outputs and the delay involved.

This delay is useful in calculating the delay involved in an N-digit BCD adder.The delays involved in the other papers [13–17] are calculated on the similar lines. This delay analysis is only approximate values as there are different types of gates are used in the various papers.

The analyses of various implementations discussed are tabulated in Table-1.It gives the comparisons of the different designs in terms of the important design parameters like number of reversible gates, number of garbage outputs, and number of constant inputs in addition to the delay parameter. From the table it is observed that the present proposal uses least number of gates producing least number of garbage outputs and has the minimum gate delay compared to the other design methods.

TABLE 1
COMPARATIVE ANALYSIS OF VARIOUS BCD ADDER IMPLEMENTATIONS [13-17]

| Reversible BCD adder | Adder-1 | | Correction circuit + Fan-out | | Adder-2 | | Complete circuit | | | Total delay to generate the sum in terms of number of gates |
|---|---|---|---|---|---|---|---|---|---|---|
| | No. of gates | No. of garbage outputs | No. of gates | No. of garbage outputs | No. of gates | No. of garbage outputs | No. of gates | No. of garbage outputs | No. of constant inputs | |
| BCD adder[13] With out Fan-out | 4 | 8 | 3 | 6 | 4 | 8 | 11 | 22 | 11 | 10 |
| BCD adder[14] | 4 | 8 | 3+3 | 6 | 4 | 8 | 14 | 22 | 17 | 13 |
| BCD adder[15] | 8 | 8 | 7 | 6 | 8 | 8 | 23 | 22 | 17 | 14 |
| BCD adder[16] | 4 | 8 | 3 | 1 | 3 | 2 | 10 | 11 | 7 | 10 |
| BCD adder[17] | 4 | 8 | 2 | 1 | 3 | 2 | 9 | 11 | 7 | 9 |
| **Proposed BCD adder** | **4** | **8** | **1** | **0** | **3** | **2** | **8** | **10** | **6** | **8** |

Because of these optimization parameters the overall cost of the circuit will be reduced. The design method definitely useful for the construction of future computer and other computational structures. Alternate optimization methods are under investigation as a future work.

## ACKNOWLEDGMENT

The authors wish to thank the ECE Department of BMS College of Engineering, Bangalore, Karnataka, India, for supporting this work.

**H.R.Bhagyalakshmi** received her B.E.degree in Electronics and communication Engineeering from Bangalore University, Karnataka, India, in 1985. Later on obtained the ME degree in Electronics in 1995 from Bangalore university, Karnataka.Currently she is an assistant professor in the department of Electronics and communication B.M.S College of engineering,Bangalore. She is pursuing her Ph.D in the Visveswaraya technological university, Begaum.Her reaerch interests include Digital circuits and logic design, reversible logic and sythesis, multiple valued logic, advanced computing techniques, low power VLSI.

**M.K.Venkatesha** received his B.E degree in Electronics and communication engineering from university of Mysore, Karnataka, India. He completed his Post Graduation from the University of Manitoba, Canada, on a Manitoba Hydro fellowship. He is a J C Bose Gold Medalist.He obtained his Ph.D in 1989 from university of Mysore, Karnataka.Currently he is heading R.N.S Institute of Technology, Bangalore, Karnataka, India. His reserach interests include digital signal processing, power electronics, digital circuits and logic design, reversible logic.

.